\documentclass[11pt,twoside]{article}
\usepackage{newpasp}
\usepackage{epsfig}

\newcommand{\xtegsfc}{XTE~J1751-305}
\newcommand{\xtemit}{XTE~J0929-314}
\newcommand{\sax}{SAX~J1808.4$-$3658}

\begin{document}
\title{Arresting Accretion Torques with Gravitational Radiation}
\author{Lars Bildsten}
\affil{Kavli Institute for Theoretical Physics and Department of
Physics, Kohn Hall, University of California, Santa Barbara, CA 93106;
bildsten@kitp.ucsb.edu}

\begin{abstract}

 Recent theoretical work has made it plausible for neutron stars (NSs)
to lose angular momentum via gravitational
radiation on long timescales ($\sim10^6$ yr) while actively accreting.
The gravitational waves (GWs) can either be emitted via the excitation
of r-modes or from a deformed crust. GW
emission can thus intervene to slow-down or halt the otherwise
relentless spin-up from accretion. Prior to this theoretical
work (and the measurements of NS rotation rates in LMXBs, see
Chakrabarty's contribution) the community was rather confident that an
accreting NS would be spun-up to rotation rates near breakup,
motivating searches for sub-millisecond objects. After only briefly
describing the physics of the GW processes, I argue that the limiting
spin frequency might be appreciably lower than the breakup
frequency. Millisecond radio pulsar observers would likely discover
the impact of GW emission as a dropoff in the number of pulsars beyond
600 Hz, and I show here that the millisecond pulsar inventory in 47
Tuc might already exhibit such a cutoff. These theoretical ideas will
be tested by GW searches with ground-based interferometers, such as
the advanced LIGO instrument proposed for operation by 2008.
\end{abstract}

\section{Introduction} 

 It has been over twenty years since the discovery of the first
millisecond radio pulsar (MSP) by Backer et al. (1982). Since then,
large numbers of MSPs have been found in globular clusters and in the
field. The theoretical expectation is that these NSs
are ``recycled'' and spun-up by prolonged accretion (see
reviews by Phinney \& Kulkarni 1994; Bhattacharya 1995).  If accreting
material arrives with the specific angular momentum of a
particle orbiting at the NS radius ($R=10 R_6 {\rm km}$), then it
takes $\approx 10^8 \ {\rm yr}$ of accretion at a rate $\dot M\approx 10^{-9}
M_\odot \ {\rm yr^{-1}}$ for a $M=1.4M_{1.4} M_\odot$ NS to reach
$\nu_s=500$ Hz from an initially low frequency (see Cook et al. 1994
for a thorough set of numbers).  Many binary scenarios (e.g. Burderi
et al. 1999) easily provide such an $\dot M$ and transfer 
enough material to spin-up the NS to submillisecond
periods. Hence, {\it if there is no limiting physics}, rapid rotation
is expected and there should be no ``cutoff'' in the population other
than that from the transferred mass.

  The piece of limiting physics typically discussed is 
magnetic accretion (e.g. Ghosh \& Lamb 1979).  For a given
$\dot M$ and magnetic dipole strength, $\mu$, the NS has a
magnetospheric radius inside of which the flow is fixed by the
magnetic topology. However, as the NS is spun-up, material rotating in
the disk at the magnetospheric radius eventually co-rotates with the
NS, such that further accretion will not likely spin it up.  Namely,
an equilibrium spin period is reached that matches the Kepler period
at the magnetosphere
\begin{equation} 
P_{eq}=8\ {\rm ms}\left({10^{-9} M_\odot \ {\rm yr^{-1}}}\over \dot
M\right)^{3/7}\left(\mu\over 10^{27} {\rm G \ cm^{3}}\right)^{6/7}, 
\end{equation} 
and further accretion is likely accomodated by alternating epochs 
of spin-up and spin-down. Such phenomena have 
been seen in the highly magnetic ($\mu>10^{30} \ {\rm G \ cm^{3}}$, 
$B>10^{12} \ {\rm G}$) accreting
pulsars in high mass X-ray binaries (see Bildsten et al. 1997 for an
overview). The $P_{eq}$'s for these NSs range from 1-1000
seconds for $\dot M\approx
10^{-10}-10^{-7} M_\odot \ {\rm yr^{-1}}$, 
clearly implying a range of $\mu$'s. 
 
\begin{table}[htb]
\begin{center}
\caption{Oscillations During Type I X-Ray Bursts}
\begin{tabular}{llllll}
\hline 
Object Name & $\nu_B$ & Flux & $h_c$ &\\ 
  & (Hz) & $10^{-9}$(cgs) & ($10^{-27}$) & \\
\hline
4U 1916-053 & 270 & 0.5 & 1.0 \\
4U 1702-429 & 330 & 1.0 & 1.2 \\
4U 1728-34 & 363 & 2.8 & 2.0 \\
KS 1731-260 & 524& 0.2-2 &1.3 \\
Aql X-1 & 549 & 1.0 & 1.0 \\
MXB 1658-298 & 567 & 0.1& 0.3\\
4U 1636-53 & 581 & 4.4 & 2.0\\
MXB 1743-29 & 589 & ? &?\\
SAXJ1750.8-2980 & 601 &? &? \\
4U 1608-52 & 619 & $<1$ & $<1$ \\
\hline
\end{tabular}
\end{center}
\end{table}

As discussed by Chakrabarty, the $\nu_s$'s inferred from nearly
coherent oscillations (given by $\nu_B$ in Table 1) during Type I
bursts finally revealed rapidly rotating NSs in LMXBs. In addition,
there are three accreting millisecond pulsars: \sax \ ($\nu_s=401 \
{\rm Hz}$; Wijnands \& van der Klis 1998), \xtegsfc \ ($\nu_s=435 \
{\rm Hz}$; Markwardt et al. 2002) and \xtemit \ ($\nu_s=185 \ {\rm
Hz}$; Galloway et al. 2002) which have $\sim 10^{-11} M_\odot \ {\rm
yr^{-1}}$. All of the $\nu_s$'s are at least a factor of two away from
the breakup frequency. It is also striking that, despite $\dot M$
contrasts of nearly a factor of 1000, the NS rotation rates are all
within a factor of 2-3 of each other. White and Zhang (1997) (and
others) have argued that this similarity arises because these NSs are
magnetic and have reached $P_{eq}$. This has two requirements: (i) a
tight relation between $\mu$ and $\dot M$ so that they all reach the
same equilibrium, and (ii) a method of hiding the persistent pulse
typically seen from a magnetic accretor.  The state of knowledge of
the $\mu$ values of these accreting NSs is quite limited (see
Cumming's contribution), though we desire them to have adequate
$\mu$'s to become MSPs.

 The current $\nu_s$ measurements of these accreting NSs thus
suggests a limiting frequency that is not easily accomodated by
either magnetic limitations or lifetime arguments. This indication,
combined with the theoretical insights summarized next,
points to GW emission as a new piece of physics
that can limit the spin-up. This would be ``bad news'' for those
carrying out searches for previously hidden large populations of
submillisecond MSPs, but ``good news'' for ground-based gravitational
wave searches.

\section{GW Emission from Crustal Quadrupoles and/or R-Modes}

Let's start with a simple hypothesis, which is that the NS has a
misaligned quadrupole moment, $Q$, induced by accretion. The strong $\nu_s$
dependence of GW emission  defines a critical frequency beyond
which accretion can no longer spin-up the star. The NS will radiate
energy via GW's at a rate $\dot E=32 GQ^2\omega^6/5 c^5$, where
$\omega=2\pi \nu_s$, and lose angular momentum at the rate
$N_{gw}=\dot E/\omega$. Balancing this spin-down torque with the
characteristic spin-up accretion torque,
$N_a\approx \dot M(GMR)^{1/2}$, gives the $Q$ needed so as to halt
spin-up (Bildsten 1998)
\begin{equation}\label{eq:qneed} 
Q\approx 10^{37} \ {\rm g \ cm^2}\left(\dot M\over 10^{-9}
\ {\rm M_\odot \ yr^{-1}}\right)^{1/2}\left(500 \ {\rm Hz}\over
\nu_s\right)^{5/2},
\end{equation} 
or $\approx 10^{-8}$ of the NS moment of inertia.
The similarities in $\nu_s$ then arise because of the
weak dependencies of the critical frequency, $\nu_{s,crit}\propto 
\dot M^{1/5} Q^{-2/5}$. 

 Bildsten (1998) and Ushomirsky et al. (2000) described physical
mechanisms in the deep NS crust that map temperature perturbations
into density perturbations in order to create a crustal quadrupole
while accreting. Quadrupolar symmetry is not required, it is just the
most efficient radiator. Ushomirsky et al. (2000) showed that 1-5 \%
lateral temperature perturbations are adequate, and that if the
composition or heating asymmetries are independent of $\dot M$, then
the induced quadrupole scales $\propto \dot M^{1/2}$ so that the
limiting $\nu_s$ weakly depends on $\dot M$. The critical open
questions to the {\it ab initio} theory are: (i) What is the cause of
the required few percent temperature perturbations? and (ii) Can the
elastic readjustment of the crust be accomodated without
cracking? This question is critical for the precession of isolated
pulsars (see Link's contribution).

 Another GW emission mechanism is that offered by the newly
discovered r-mode instability for the fluid in the NS core.  In a
series of remarkable breakthroughs, the gravitational physics
community has shown (see Andersson's 2002 recent review for
references) that r-modes (those waves which get a large part of their
restoring force via the Coriolis term) are unstable to the emission of
gravitational radiation via the Chandrasekhar-Friedman-Schutz
instability. Namely, the radiation of GWs excites the modes,
regardless of spin frequency!

The beauty of the r-modes is that there is no need to invoke an
asymmetry. That part comes immediately once it has been shown that the
GW excitation beats the viscous damping. However, finding the minimum
$\nu_s$ to trigger the instability involves understanding the
viscosity of the NS core and the coupling of the r-modes to the
overlying crust (e.g. Levin \& Ushomirsky 2001; Wagoner 2002). The
critical $\nu_s$ depends on both the NS structure and core
temperature, and current calculations place the accreting NSs squarely
in the transition regime between stable and unstable. Hence, the
hypothesis of r-mode instability remains quite viable. How the angular
momentum is lost by GW emission once the mode goes unstable and starts
to grow is complicated by Levin's (1998) discovery that the internal
heating from a runaway r-mode leads to a limit cycle behavior and
transient GW emission. The duty cycle for GW emission is
set by how large the mode amplitude can be  before non-linear
effects set in. Arras et al.'s (2002) current estimates of the mode
saturation imply that the GW emission occurs for $10^{3-4}$ years
every $\sim 100$ Myr.  Most recently, Wagoner (2002) showed that the
damping microphysics is uncertain enough that it is possible for
steady-state r-mode emission with an internal wave amplitude of $\sim
10$ cm.

\begin{figure}
\begin{center}

  \includegraphics[width=0.63\textwidth]{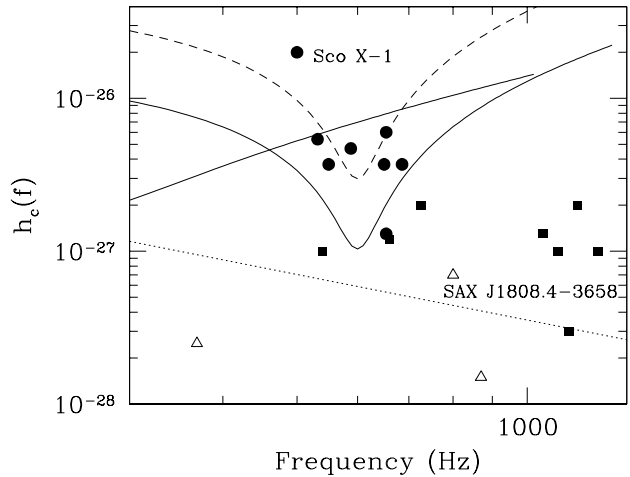}
    \caption{Periodic GW strains at Earth for Type I
bursters (solid squares), accreting millisecond pulsars
(open triangles) and the bright Z sources (filled circles, 
where I use the khz QPO difference frequency as $\nu_s$)
    presuming quadrupolar emission in equilibrium with accretion. 
The two solid lines are 
    proposed sensitivities for LIGO-II (courtesy K. Strain,
    1999) with and without narrowbanding for 3 month integrations. 
The dashed line is a 2 week LIGO-II search. The lower dotted
  line is LIGO-II's proposed thermal noise floor, which is the best
  case for narrowbanding at any frequency. }
\end{center}
\vspace{-.20in}
\end{figure}

  For persistent GW emission under either scenario, the GW strain at
Earth from a source at distance $d$ can be found. For simplicity, I do
it here for the quadrupolar deformation (the GW signal is then at
$2\nu_s$, for r-modes it would be $4\nu_s/3$) where $h_c=2.9 G\omega^2
Q/dc^4$ after suitably averaged over spin orientations (Brady et
al. 1998).  Presuming the NS luminosity is $L\approx GM\dot M/R$ then
$h_c$ is written in terms of the observable $F=L/4\pi d^2$ (Wagoner
1984)
\begin{equation} \label{eq:hsecond}
h_c\approx 1.3 \times 10^{-27}{R_6^{3/4}\over M_{1.4}^{1/4}}
\left(F\over 10^{-9} \ {\rm erg \ cm^{-2} \ s^{-1}}\right)^{1/2}
\left(300 \ {\rm Hz}\over \nu_s\right)^{1/2},  
\end{equation}
which represents a lower limit for $h_c$ due to the $L$ to $\dot M$
conversion I chose and since $F$ is never fully measured. The minimum
$h_c$'s for those NSs which have $\nu_s$ inferred from Type I bursts
are shown in Table 1 and plotted in Figure 1. The average 2-10 keV
fluxes were either from van Paradijs (1995) or my own estimates.  The
figure caption describes the curves (see Cutler and Thorne 2002 for
details) and makes it clear that Sco X-1 is an excellent
target. However, the simplest LIGO-II search would be one where the
pulse ephemeris is known and that is only true for the three accreting
ms pulsars.  The best is \sax \ at $\nu_s=401 $ Hz, which is nearby,
has accretion outbursts roughly every two years, and a well known
orbit (Chakrabarty \& Morgan 1998).  A dedicated search at 802 Hz with
a tuned LIGO-II could bear fruit.

\begin{figure}
\begin{center}
   \includegraphics[width=0.7\textwidth]{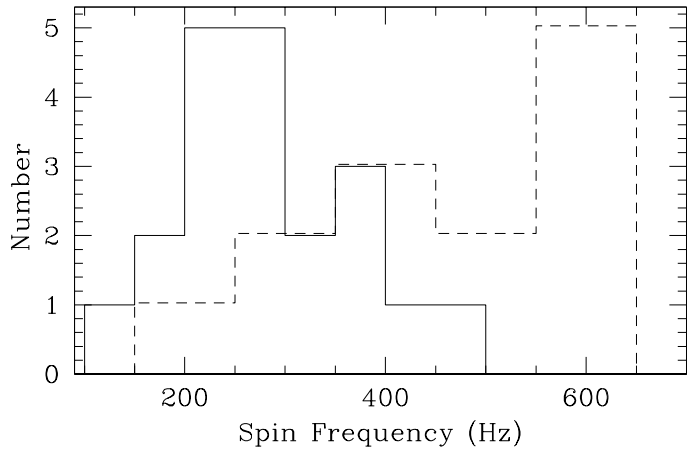}
    \caption{Histograms of NS $\nu_s$'s.  The solid
line is the $\nu_s$ distribution for MSPs in 47 Tuc (Camilo et
al. 2000), whereas the 
dashed line is for the NSs in Table 1 and the 3 
accreting millisecond pulsars.}
\end{center}
\vspace{-.20in}
\end{figure}

\section{Conclusions and Implications for Millisecond Radio Pulsar  Searches}

I have hopefully made the case that both theory and observation point
to a new limiting mechanism to NS spin-up: GW emission.  If true, the
impact on MSP searches is clear: {\it many fewer rapidly rotating MSPs
will be found than predicted in those scenarios that neglect this
physics.} One way that such physics could reveal itself is to put in a
``cutoff'' in the MSP population at some particular frequency. The
prime complication is attributing such a cutoff to the underlying
population and not observational selection.

   However, in 47 Tuc, the observational selection is calculable, and
between 300 and 600 Hz, the MSP sensitivity drops by only 30 percent
(see Figure 2 of Camilo et al. 2000). Such a small sensitivity drop is
not adequate to explain the population decline (see solid line in
Figure 2) to high frequencies in 47 Tuc. I also plot the $\nu_s$
distribution for 13 accreting NS with a dashed line. This sample is
not, to our knowledge, biased in frequency in any way. If this is the
injection distribution that becomes like that in 47 Tuc, then the
higher $\nu_s$ systems must spin-down a factor of 1.5 in 5-8 Gyrs,
easily accomodated with a surface field of $B\approx (2-5)\times 10^8
{\rm G}$. 

\noindent
\acknowledgements

I thank P. Arras, D. Chakrabarty and A. Cumming for comments and 
my previous co-authors on this work, Curt Cutler and Greg
Ushomirsky, for many insights. This research was supported by NASA via grant
NAG 5-8658 and by the NSF under Grants PHY99-07949 and
AST02-05956. L. B. is a Cottrell Scholar of the Research Corporation.

\end{document}